\documentclass[aps,twocolumn,showpacs,superscriptaddress,amsmath,amssymb,nofootinbib]{revtex4-1}

\usepackage{graphicx}
\usepackage{dcolumn}% Align table columns on decimal point
\usepackage{bm}% textbf math
\usepackage{color}
\usepackage{txfonts}
\usepackage{hyperref}
\usepackage{soul}
\usepackage{upgreek}
\usepackage{CJK}

\def\NESTED{\mbox{\begin{picture}(7,11)
\put(1,10){\line(1,0){10}}
\put(1,0){\line(1,1){10}}
\put(1,0){\line(0,1){10}}
\end{picture}
}}

\def\CPSHAPE{\mbox{\begin{picture}(7,11)
\put(1,10){\line(1,0){10}}
\put(1,0){\line(0,1){10}}
\put(1,0){\line(1,0){5}}
\put(6,0){\line(0,1){5}}
\put(6,5){\line(1,0){5}}
\put(11,5){\line(0,1){5}}
\end{picture}
}}

\begin{document}
\begin{CJK}{UTF8}{mj}
\title{Network nestedness as generalized core-periphery structures}

\author{Sang Hoon Lee (이상훈)}
\email{lshlj82@kias.re.kr}
\affiliation{School of Physics, Korea Institute for Advanced Study, Seoul 02455, Korea}

%%%%%%%%%%

\begin{abstract}
The concept of nestedness, in particular for ecological and economical networks, has been introduced as a structural characteristic of real interacting systems. We suggest that the nestedness is in fact another way to express a mesoscale network property called the core-periphery structure. With real ecological mutualistic networks and synthetic model networks, we reveal the strong correlation between the nestedness and core-periphery-ness (likeness to the core-periphery structure), by defining the network-level measures for nestedness and core-periphery-ness in the case of weighted and bipartite networks. However, at the same time, via more sophisticated null-model analysis, we also discover that the degree (the number of connected neighbors of a node) distribution poses quite severe restrictions on the possible nestedness and core-periphery-ness parameter space. Therefore, there must exist structurally interwoven properties in more fundamental levels of network formation, behind this seemingly obvious relation between nestedness and core-periphery structures.
\end{abstract}

% PACS, the Physics and Astronomy % Classification Scheme.%%
%%87.23.-n Ecology
%%89.75.Fb Structures and organization in complex systems
%%89.75.Hc Networks and genealogical trees
%%92.40.Oj Plant ecology
%\keywords{}
%Use showkeys class option if keyword %display desired

\maketitle

%%%%%%%%%%%%%

\section{Introduction}
\label{sec:introduction}

Since the pioneering work by May~\cite{May1972}, the concept of nestedness indicating the systematically included structure composed of generalists and specialists has been assumed to be one of the most characteristic structures of ecological networks~\cite{Roberts1974,AlmeidaNeto2008,Corso2008,Galeano2009,Bastolla2009,Allesina2012,Rohr2014}. These obviously evolved, not designed, systems must have reasons to be formed as such, and the candidates for the reasons include dynamical stability~\cite{May1972,Roberts1974,Allesina2012,Rohr2014} and biodiversity~\cite{Bastolla2009}. Once such a structural property is expressed as a purely mathematical form, it is possible to study the network systems in general without involving the intrinsic properties of ecosystems. Indeed, compared to when the concept was first conceived, such networked systems in general have been widely investigated since the turn of the century and so on~\cite{NetworkReview}, which naturally enables us to connect the nestedness to possibly more network properties in more general contexts. For instance, the concept has been used to describe economic systems~\cite{Saavedra2009,Saavedra2011} such as industrial ecosystems~\cite{Bustos2012} as well.

In such a general setting, nestedness is one of the examples of the \emph{mesoscale} structure of networks. The term mesoscale means somewhere between the microscale structure such as the degree (the number of neighbors a node has) and the macroscopic structure such as the average edge density (the ratio of the number of existing edges to the number of node pairs). In this paper we suggest that another mesoscale structure, called the core-periphery structure of networks~\cite{Borgatti1999,Holme2005,Csermely2013,Rombach2014,SHLee2014,Cucuringu2014}, is closely related to nestedness. To be more specific, the gradual change from generalists to specialists corresponds to the gradual change in the \emph{coreness} of the nodes, which makes the nested structure a generalized version of a clear-cut core versus periphery structure. 
In fact, the connection between the two concepts may seem to be too obvious from the adjacency matrix form (\NESTED shape for the nested and \CPSHAPE shape for the core-periphery structure) to report, but we also find that it reveals a more fundamental property of real networks constrained by the degree distribution, which is widely used as the keystone of network ensembles. 

To systematically investigate the relation between the two concepts, we first define the measures of nestedness and core-periphery-ness (likeness to the core-periphery structure) in the weighted and bipartite network level in Sec.~\ref{sec:nestedness_and_c_p}. The mutualistic ecological networks and synthetic network model used in our study are presented in Sec.~\ref{sec:mutualistic_networks}. Using the measures and data introduced, we present the result in Sec.~\ref{sec:results}. We conclude the paper in Sec.~\ref{sec:discussion} with a summary and discussion.

\section{Measures for Nestedness and Core-periphery Structures}
\label{sec:nestedness_and_c_p}

\subsection{Nestedness}
\label{sec:nestedness}

We use the basic nestedness metric based on overlap and decreasing fill (NODF)~\cite{AlmeidaNeto2008}, denoted by $\nu$ in this paper for simplicity, although we note that there are other measures~\cite{Corso2008,Galeano2009}. The NODF counts the number of pairs of rows satisfying the nested structure for each column pair and the number of columns satisfying the nested structure for each row pair, after sorting the rows and columns in descending order of degree, or strength (the sum of weights on the edges attached to the node) in the case of weighted networks~\cite{Barrat2004}. Suppose $\{ W_{ij} \}$ is the weighted adjacency matrix~\cite{NetworkReview} representing the network ($W_{ij} > 0$ represents the interaction between nodes $i$ and $j$, while $W_{ij} = 0$ represents the absence of interaction between $i$ and $j$), where both sets of indices are sorted by the descending order of nodes' strength. Mathematically,
\begin{equation}
\nu = \frac{\displaystyle \sum_{r=1}^{n(n-1)/2} f_{\mathrm{col}}(r) + \sum_{c=1}^{m(m-1)/2} f_{\mathrm{row}}(c)}{n(n-1)/2 + m(m-1)/2} \,,
\label{eq:NODF}
\end{equation}
where $f_{\mathrm{col}}(r) \in [0,1]$ and $f_{\mathrm{row}}(c) \in [0,1]$ are the fraction of pairs of columns satisfying the nested inclusion structure for the row pair index $r$ and the fraction of pairs of rows satisfying the nested inclusion structure for the column pair index $c$, respectively, for the adjacency matrix $\{ W_{ij} \}$, and $n$ and $m$ are the numbers of rows and columns, respectively, which are used in the denominator for the proper normalization condition $\nu \in [0,1]$. As a result, $\nu$ captures the maximally nested case ($\nu = 1$) and the minimally nested case ($\nu = 0$). For details with illustrations, see Ref.~\cite{AlmeidaNeto2008}.

Note that we generalize the inclusion criterion introduced in Ref.~\cite{AlmeidaNeto2008} to include the weighted networks, as many of our mutualistic networks (introduced in Sec.~\ref{sec:data_sets}) are weighted. The generalization is straightforwardly achieved by using the unviolated-case criterion that contributes to $f_\mathrm{col}(r)$: $W_{ij} \le W_{ij'}$ for $j < j'$ (the strength of $j$ is greater than or equal to the strength of $j'$) instead of the unweighted version that $W_{ij}$ should be $0$ if $W_{i'j}$ is $0$ for $j < j'$. The unviolated case contributing to $f_\mathrm{row}(c)$ is similar: $W_{ij} \le W_{i'j}$ for $i < i'$ (the strength of $i$ is greater than or equal to the strength of $i'$) instead of the unweighted version that $W_{ij}$ should be $0$ if $W_{i'j}$ is $0$ for $i < i'$. In the case of unweighted networks where $W_{ij} \in \{0,1\}$, our criterion is just the same as the conventional one for the unweighted version in Ref.~\cite{AlmeidaNeto2008}, which is used for our synthetic networks (introduced in Sec.~\ref{sec:synthetic_network_model}).

\subsection{Core-periphery Structures}
\label{sec:core_periphery}

One may argue that the nestedness and core-periphery structure are different in spirit, as the former describes the overall organization of a network and the latter focuses on the separation of core and periphery. However, as illustrated in Refs.~\cite{Rombach2014,SHLee2014,Cucuringu2014}, the latter also concerns the overall structures by assigning the \emph{gradual} core scores for nodes (and edges as well; see Refs.~\cite{SHLee2014,Cucuringu2014} for details). As we demonstrate in Sec.~\ref{sec:results}, the nested structures of adjacency matrices are in fact such a gradual change of coreness. Of course, one can always define certain objective functions analogous to the community identification~\cite{CommunityReview} to actually find the core-periphery separation, e.g., as presented in Ref.~\cite{Cucuringu2014}.

The method to calculate the edge-density-based coreness, called a core score (CS), is the modified version of the one introduced in Refs.~\cite{Rombach2014,SHLee2014,Cucuringu2014} to fully consider the bipartivity of ecological mutualistic networks. There are other ways to quantify the coreness such as backup-path-based one in Refs.~\cite{SHLee2014,Cucuringu2014}, but we use CS in this analysis because there is a natural way to quantify the overall core-periphery structure in the formalism of CS. Again, suppose $\{ W_{ij} \}$ is the weighted adjacency matrix representing the mutualistic network, in particular, among a given set of animals $\{1,\ldots , N_\mathrm{animal}\}$ and plants $\{1,\ldots , N_\mathrm{plant}\}$. The network has $N = N_\mathrm{animal} + N_\mathrm{plant}$ nodes in total, and the value $W_{ij}$ indicates the weight of the connection between the animal node $i \in \{1, \ldots , N_\mathrm{animal}\}$ and the plant node $j \in \{1, \ldots , N_\mathrm{plant}\}$. We insert the core-matrix elements $\{ \mathsf{C}_{ij} \}$ into the core quality
\begin{equation}
	R({{\upalpha},{\upbeta}}) = \sum_{i,j} W_{ij} \mathsf{C}_{ij}({\upalpha},{\upbeta})\,, \label{core_quality}
\end{equation}
where the components of the parameter vector ${\upalpha} = (\alpha_\mathrm{animal},\alpha_\mathrm{plant}) \in [0,1]$ determines the sharpness of the core-periphery division and ${\upbeta} = (\beta_\mathrm{animal},\beta_\mathrm{plant}) \in [0,1]$ determines the fraction of
core nodes for animals and plants, respectively. We decompose the core-matrix elements into a product form, $\mathsf{C}_{ij}({\upalpha},{\upbeta}) = C_i(\alpha_\mathrm{animal},\beta_\mathrm{animal}) C_j(\alpha_\mathrm{plant},\beta_\mathrm{plant})$, where the elements of the core vector
\begin{widetext}
\begin{equation}
	C_i (\alpha_\omega,\beta_\omega) = \begin{cases}
\frac{\displaystyle i(1-\alpha_\omega)}{\displaystyle 2 \lfloor \beta_\omega N_\omega \rfloor} & \mathrm{for} ~ i \le \lfloor \beta_\omega N_\omega \rfloor , \\
\\
\frac{\displaystyle (i-\lfloor \beta_\omega N_\omega \rfloor)(1-\alpha_\omega)}{\displaystyle 2(N_\omega-\lfloor \beta_\omega N_\omega \rfloor)} + \frac{\displaystyle 1+\alpha_\omega}{\displaystyle 2} & \mathrm{for} ~ i > \lfloor \beta_\omega N_\omega \rfloor \,,
\end{cases}
\label{transition_function}
\end{equation}
\end{widetext}
for each type of node $\omega \in (\mathrm{animal},\mathrm{plant})$. References~\cite{Rombach2014,SHLee2014} also discuss the use of alternative transition functions to the one in Eq.~(\ref{transition_function}).

We wish to determine the core-vector elements in (\ref{transition_function}) so that the core quality in Eq.~(\ref{core_quality}) is maximized.  This yields a CS value denoted by $\Xi$ for node $i$ of
\begin{equation}
\Xi_{\omega} (i) = Z_{\omega} \sum_{({\upalpha},{\upbeta})} C_i ({\upalpha},{\upbeta}) R({\upalpha},{\upbeta})\,,
\label{CS_formula}
\end{equation}
where the normalization factor $Z_{\omega}$ is determined so that the maximum value of $\Xi$ over the entire set of nodes is 1 for each type $\omega$ of node separately as $\omega \in \{\mathrm{animal},\mathrm{plant}\}$. As in Refs.~\cite{Rombach2014,SHLee2014}, we use simulated annealing~\cite{Kirkpatrick1983} (with the same cooling schedule as in that paper). The core-quality landscape tends to be less sensitive to $\upalpha$ than it is to $\upbeta$; for computational tractability, we fix the value of $\alpha_\mathrm{animal} = \alpha_\mathrm{plant} = 1/2$ and take the same value of  $\Delta \beta_\mathrm{animal} = \Delta \beta_\mathrm{plant} = 0.01$ and thus consider $101^2$ evenly spaced points in the ${\upbeta} = (\beta_\mathrm{animal},\beta_\mathrm{plant})$ plane.
Finally, to define the coreness of an entire network, we define
the normalized core quality inspired by Refs.~\cite{Rombach2014,SHLee2014}, denoted by $\xi$ in this paper, as
\begin{equation}
\xi = \frac{\displaystyle \sum_{i,j} W_{ij} \Xi_\mathrm{animal} (i) \Xi_\mathrm{plant} (j)}{\displaystyle \sum_{i,j} W_{ij} \sum_{i,j} \Xi_\mathrm{animal} (i) \Xi_\mathrm{plant} (j)} \, ,
\label{NCQ_formula}
\end{equation}
for the animal nodes $i \in \{1, \ldots , N_\mathrm{animal}\}$ and the plant nodes $j \in \{1, \ldots , N_\mathrm{plant}\}$,
which we use for the coreness measure throughout this paper.

\section{Data and Synthetic Model}
\label{sec:mutualistic_networks}

\subsection{Ecological Network Data}
\label{sec:data_sets}

\begin{figure}
\includegraphics[width=\columnwidth]{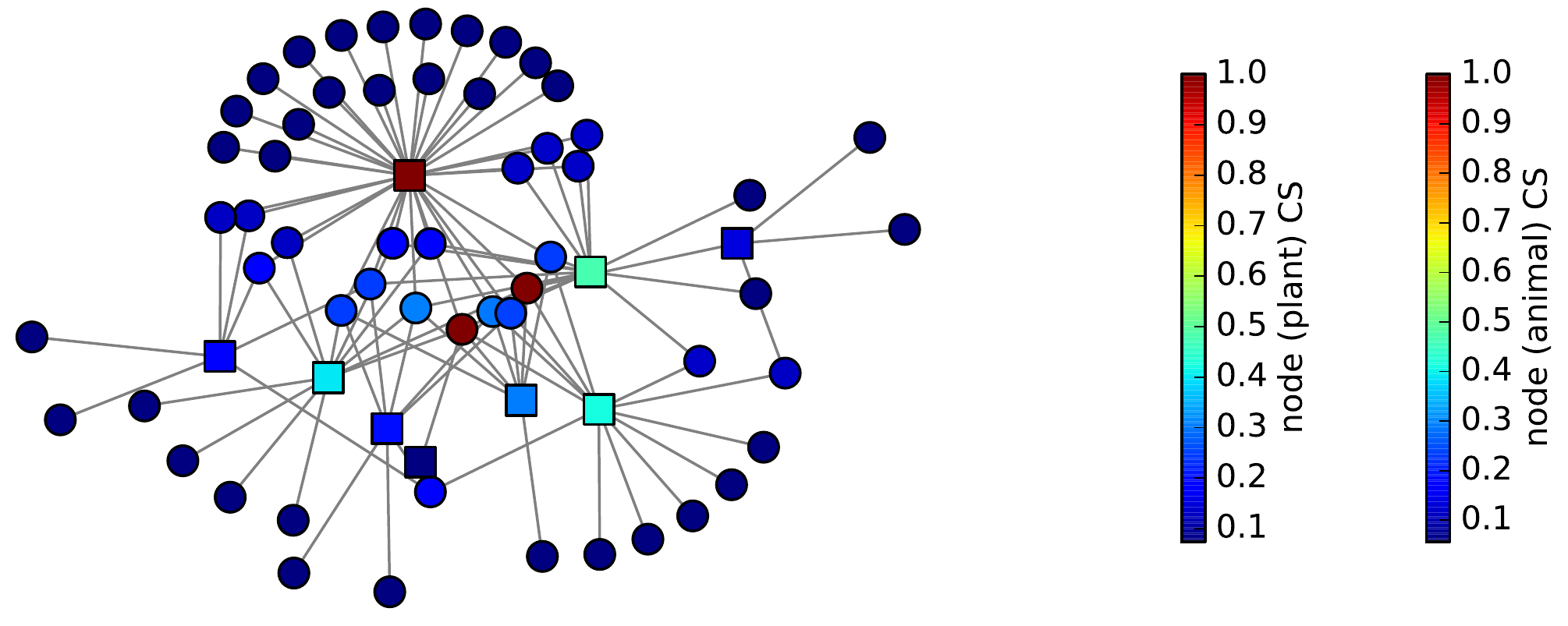} 
\caption{Mutualistic pollinator-plant network observed in the KwaZulu-Natal region, South Africa~\cite{Rohr2014,WebOfLife},
 where the $\Xi$ values in Eq.~\eqref{CS_formula} for nodes~\cite{Rombach2014,SHLee2014,Cucuringu2014} are colored.
Animals (pollinators) and plants are represented as circles and squares, respectively.
The Kamada-Kawai graph layout algorithm~\cite{Kamada1989} is used.
}
\label{KwaZulu_Natal}
\end{figure}

\begin{figure*}
\begin{tabular}{lll}
(a) & (b) & (c) \\
\includegraphics[width=0.33\textwidth]{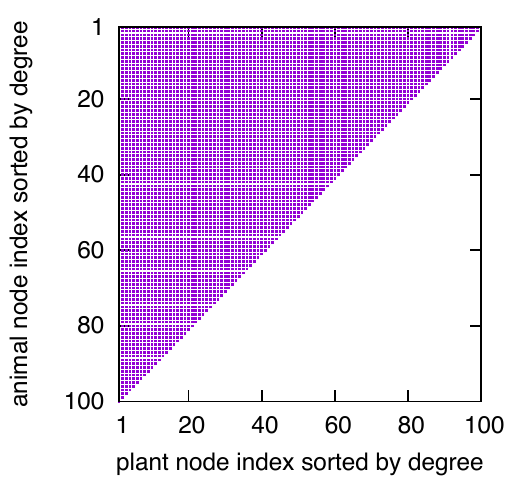} &
\includegraphics[width=0.33\textwidth]{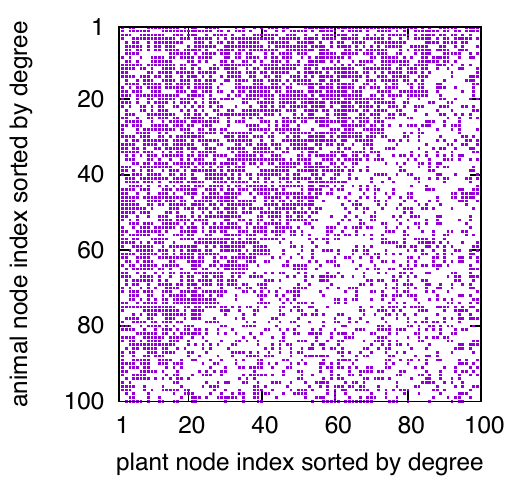} & 
\includegraphics[width=0.33\textwidth]{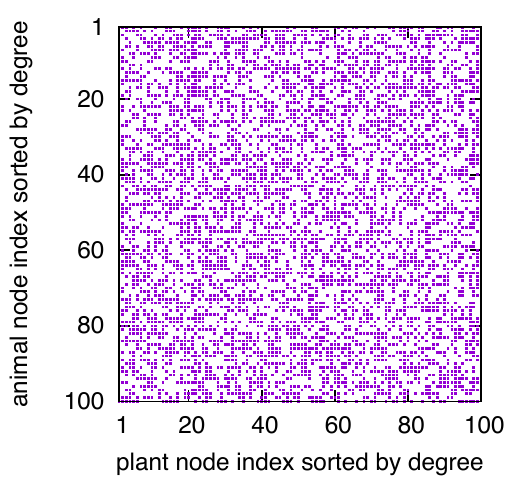} \\
\end{tabular}
\caption{Examples of adjacency matrices for our synthetic network model with different
noise parameter $\eta$, where the purple squares and empty sites represent $1$ and $0$, respectively: (a) $\eta = 0$ (a perfectly nested structure), (b) $\eta = 1/2$, and (c) $\eta = 1$.
}
\label{synthetic_adj_mat}
\end{figure*}

We use 89 mutualistic network data introduced in Ref.~\cite{Rohr2014} that can be downloaded in Ref.~\cite{WebOfLife}, consisting of 59 pollination networks and 30 seed dispersal networks.
Some networks are weighted by the interaction strength, while the others are unweighted.
The size of networks varies greatly, from the smallest one composed of 6 nodes (3 animals and 3 plants) to the largest one composed of 997 nodes (883 animals and 114 plants). Such size diversity provides us with a nice opportunity to cross-check the correlation between various measures for the system size varying across the two orders of magnitude.
Figure~\ref{KwaZulu_Natal} shows one example of a network with the core score~\cite{Rombach2014,SHLee2014,Cucuringu2014} defined in Eq.~\eqref{CS_formula}.

\subsection{Synthetic Network Model}
\label{sec:synthetic_network_model}

To control the various effects of other properties of real networks that will be discussed in Sec.~\ref{sec:results},
we construct the series of synthetic unweighted networks with tunable nestedness. 
First, we construct the perfectly nested structure shown in Fig.~\ref{synthetic_adj_mat}(a) with given numbers of
animal and plants. One can see the similarity between these nested structures represented in the adjacency matrix \NESTED~and the core-periphery structure \CPSHAPE~as shown in Fig.~1.1(b) in Ref.~\cite{Rombach2014}. In Ref.~\cite{Cucuringu2014}, the possibility of generalization of such a step structure is presented, and the finest step structure would be the perfectly nested structure in Fig.~\ref{synthetic_adj_mat}(a), indeed. In this respect, we regard the nested structure as a generalized core-periphery structure.
Starting from this perfectly nested structure, we add noise with a certain probability $\eta$, i.e., for each existing edge, the edge is removed, and a randomly chosen node pair that is currently not connected
is connected with probability $\eta$. Figure~\ref{synthetic_adj_mat} shows some examples with various $\eta$ values for $100$ animals and $100$ plants.
However, even this model does not preserve the degree sequence (thus the effect of degree heterogeneity), which yields nontrivial correlations in regard to degree,
as presented in Sec.~\ref{sec:results}.

\begin{figure}
\includegraphics[width=\columnwidth]{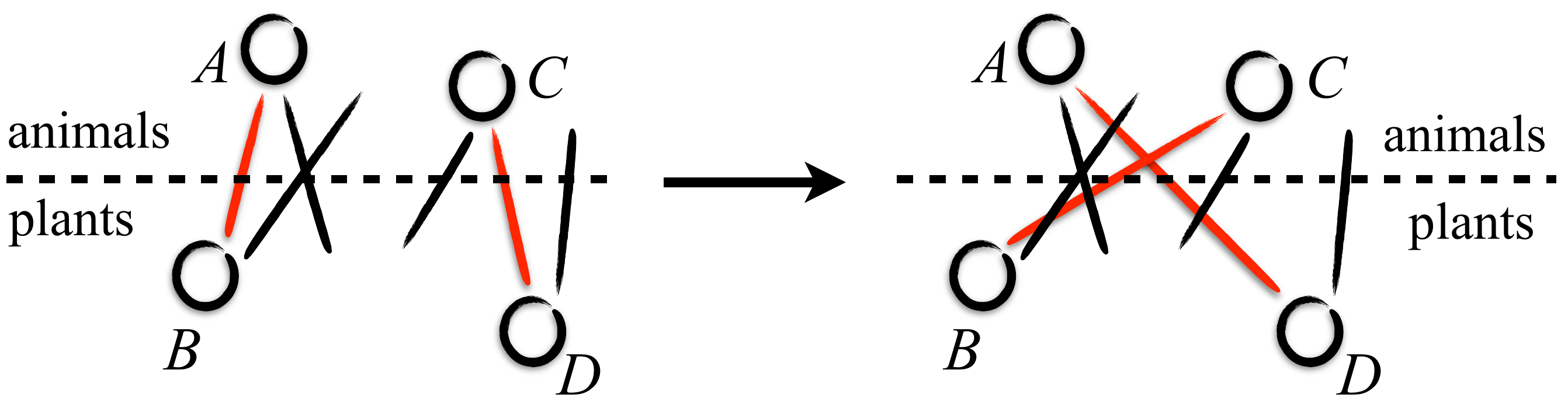} 
\caption{Illustration of the EPS method, where the pair of edges (the edges in red) $A$--$B$ and $C$--$D$ are switched to $A$--$D$ and $B$--$C$.
}
\label{edge_pair_shuffling_illustration}
\end{figure}

\begin{figure*}
\begin{tabular}{lll}
(a) & (b) & (c) \\
\includegraphics[width=0.33\textwidth]{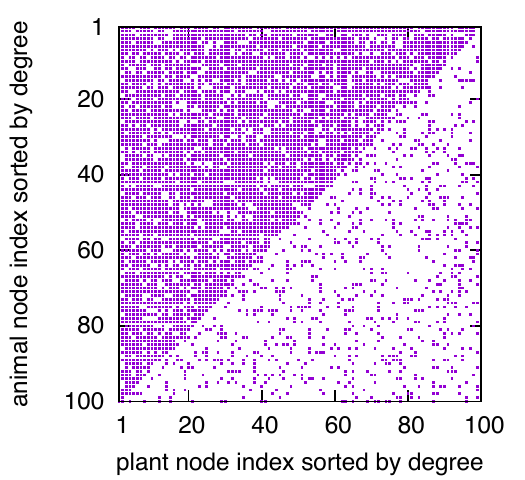} &
\includegraphics[width=0.33\textwidth]{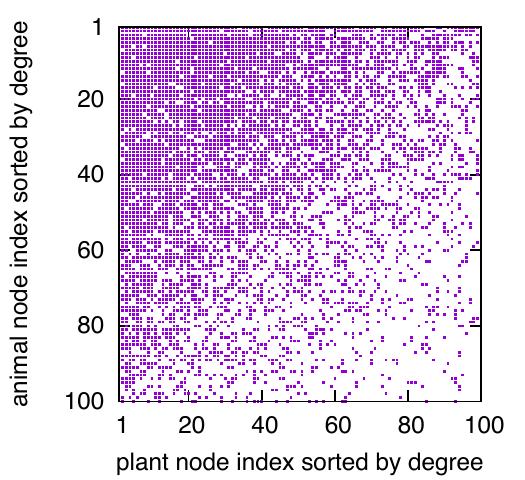} & 
\includegraphics[width=0.33\textwidth]{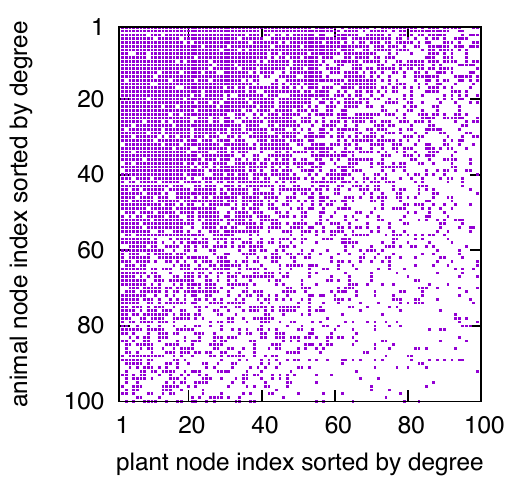} \\
\end{tabular}
\caption{Examples of adjacency matrices for the EPS method illustrated in Fig.~\ref{edge_pair_shuffling_illustration} with different
shuffling step length $T$ in the unit of the number of edges, starting from our synthetic network model with $\eta=0.2$: (a) $T=0$ (an original synthetic network with $\eta=1/5$), 
(b) $T=1/2$, and (c) $T=1$. The purple squares and empty sites represent $1$ and $0$, respectively.
}
\label{EPS_synthetic_adj_mat}
\end{figure*}

To get rid of any effect from the degree distribution or sequence, for a given network structure, 
we present an additional randomization scheme called the edge-pair-shuffling (EPS) process, illustrated in Fig.~\ref{edge_pair_shuffling_illustration}.
Since there does not exist a possible pair for swapping in the perfectly nested structure illustrated in Fig.~\ref{synthetic_adj_mat}(a)
(suppose that $A$ and $B$ are more ``generalist'' than $C$ and $D$, without loss of generality, then there should always be the edges $A$--$D$ and $B$--$C$ in that case),
we start from our synthetic network model with given $\eta$ values and apply the EPS process for selected edge pairs ($A$--$B$ and $C$--$D$, 
when both $A$--$D$ and $B$--$C$ do not exist, in Fig.~\ref{edge_pair_shuffling_illustration}) uniformly at random
repeatedly $T$ Monte Carlo steps in the unit of the number of edges.
Figure~\ref{EPS_synthetic_adj_mat} shows some examples with various $T$ values for the $100$ animals and $100$ plants.
Note that the EPS process cannot destruct the nested structure, because 
there is a fundamental constraint of graphicality for a given degree sequence in a bipartite network~\cite{GaleRyser}.
In fact, it is quite the opposite.
Somewhat counterintuitively, the average $\nu$ value is slightly \emph{increased} as we increase the number of Monte Carlo steps $T$, as presented in Sec.~\ref{sec:results}.

\section{Results}
\label{sec:results}

\begin{figure*}
\begin{tabular}{lll}
(a) & (b) & (c) \\
\includegraphics[width=0.33\textwidth]{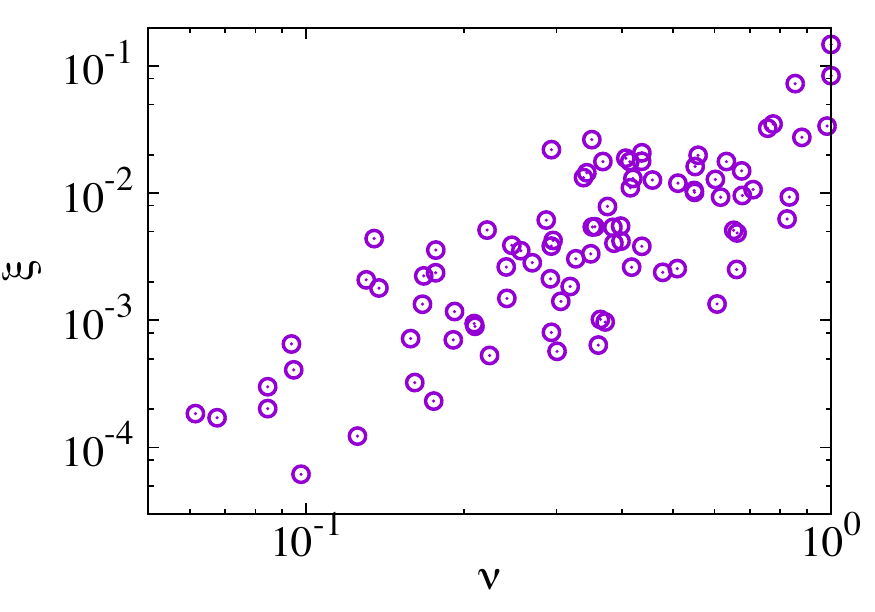} &
\includegraphics[width=0.33\textwidth]{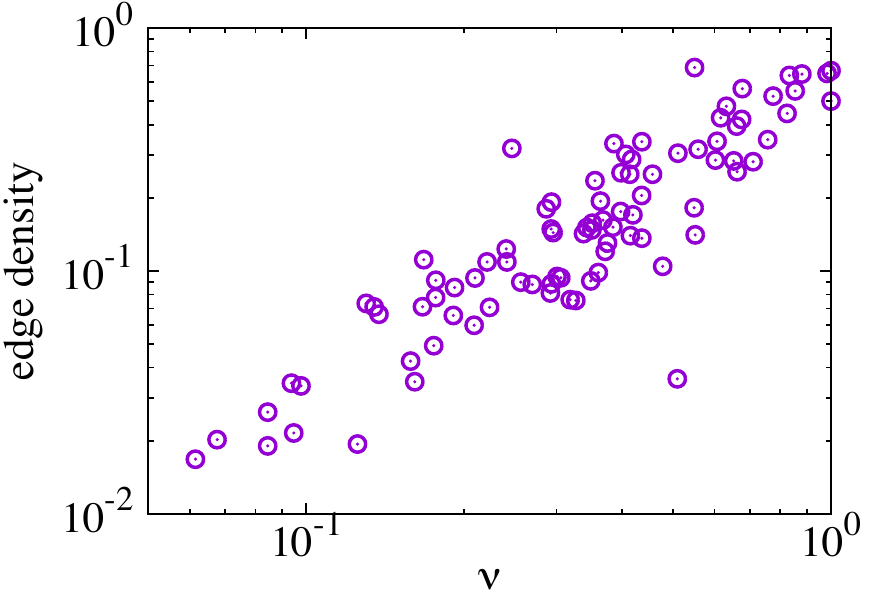} &
\includegraphics[width=0.33\textwidth]{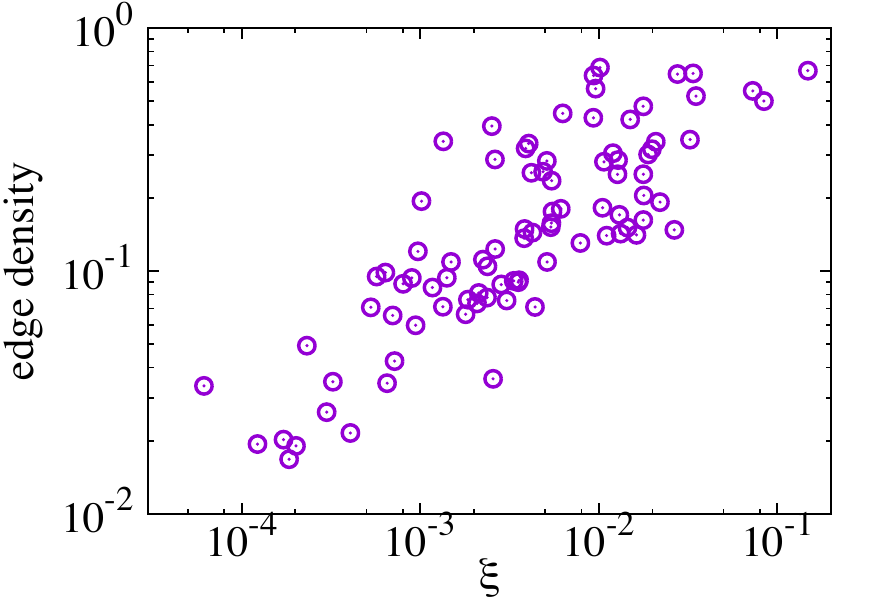} \\
\end{tabular}
\caption{For the mutualistic network data, (a) $\nu$ in Eq.~\eqref{eq:NODF} versus $\xi$ in Eq.~\eqref{NCQ_formula}, with the
correlation coefficients $0.634$ with the $p$-value less than $10^{-10}$ (Pearson), $0.751$ with the $p$-value less than $10^{-16}$ (Spearman), and $0.559$ with the $p$-value less than $10^{-14}$ (Kendall),
(b) $\nu$ in Eq.~\eqref{eq:NODF} versus the edge density, with the
correlation coefficients $0.875$ with the $p$-value lee than $10^{-28}$ (Pearson), $0.878$ with the $p$-value less than $10^{-28}$ (Spearman), and $0.721$ with the $p$-value less than $10^{-22}$ (Kendall), and
(c) $\xi$ in Eq.~\eqref{NCQ_formula} versus
the edge density, with
the
correlation coefficients $0.595$ with the $p$-value less than $10^{-9}$ (Pearson), $0.798$ with the $p$-value less than $10^{-20}$ (Spearman), and $0.614$ with the $p$-value less than $10^{-18}$ (Kendall).
}
\label{NODF_vs_NCQ_scatter}
\end{figure*}

\begin{figure*}
\begin{tabular}{lll}
(a) & (b) & (c) \\
\includegraphics[width=0.33\textwidth]{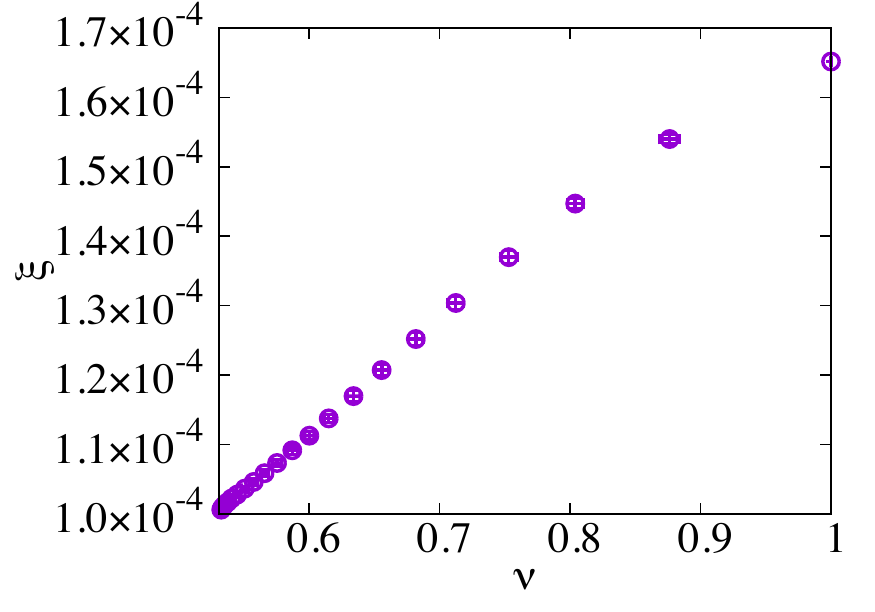} &
\includegraphics[width=0.33\textwidth]{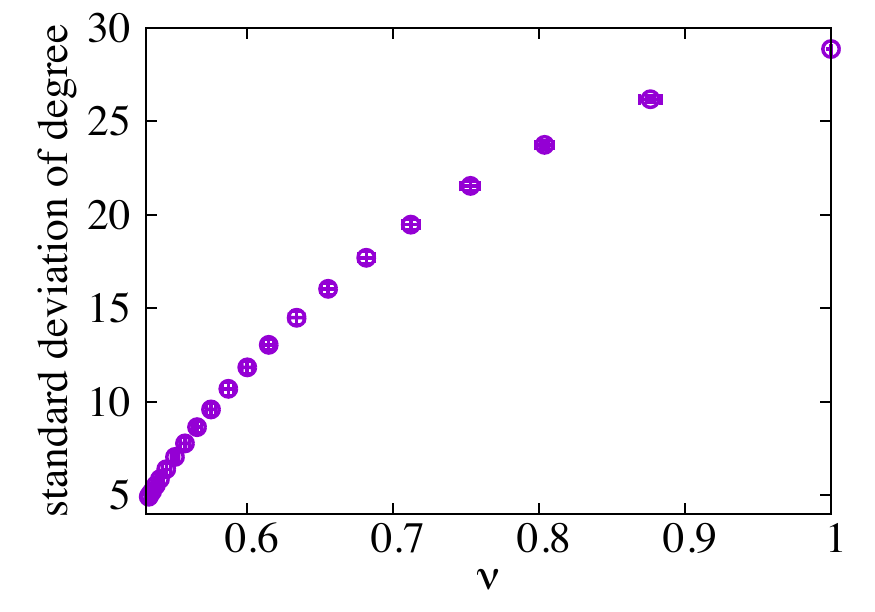} &
\includegraphics[width=0.33\textwidth]{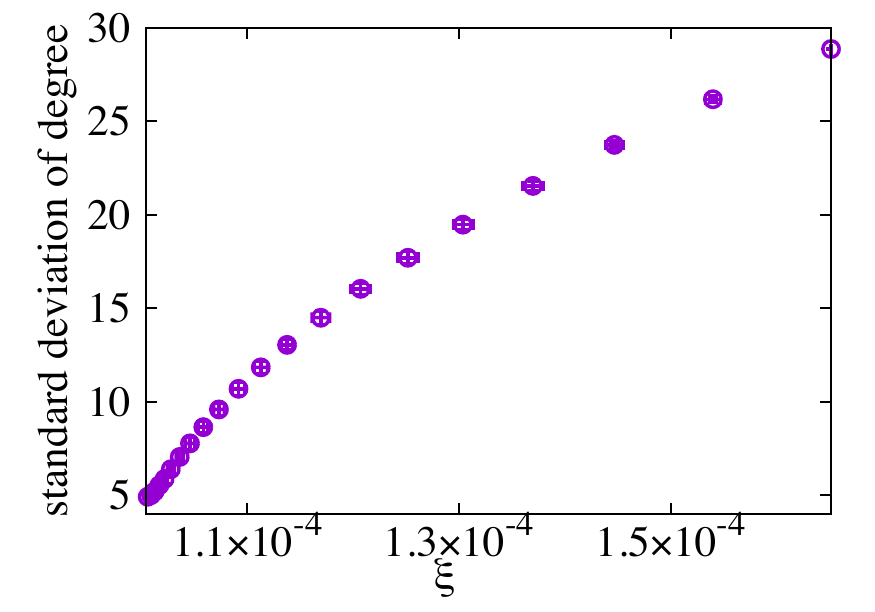} \\
\end{tabular}
\caption{For the synthetic network model with $100$ animals and $100$ plants, and the noise parameter $\eta$, as shown in Fig.~\ref{synthetic_adj_mat}, (a) $\nu$ in Eq.~\eqref{eq:NODF} versus $\xi$ in Eq.~\eqref{NCQ_formula},
(b) $\nu$ in Eq.~\eqref{eq:NODF} versus the standard deviation of degree, and 
(c) $\xi$ in Eq.~\eqref{NCQ_formula} versus the standard deviation of degree.
For all panels, the points from the lower (left) correspond to $\eta = 1, 0.95, 0.9, 0.85, \ldots, 0.1, 0.05$, and $0$
[perfect nestedness: Fig.~\ref{synthetic_adj_mat}(a)]. The error bars, most of which are smaller than the symbol size, in both the horizontal and vertical directions present the standard deviation
for each $\eta$ (100 network realizations for each $\eta$), except for $\eta = 0$ (undefined because there is only one possible realization).
}
\label{NODF_vs_NCQ_scatter_for_synthetic_networks}
\end{figure*}

\begin{figure*}
\begin{tabular}{ll}
(a) & (b) \\
\includegraphics[width=0.4\textwidth]{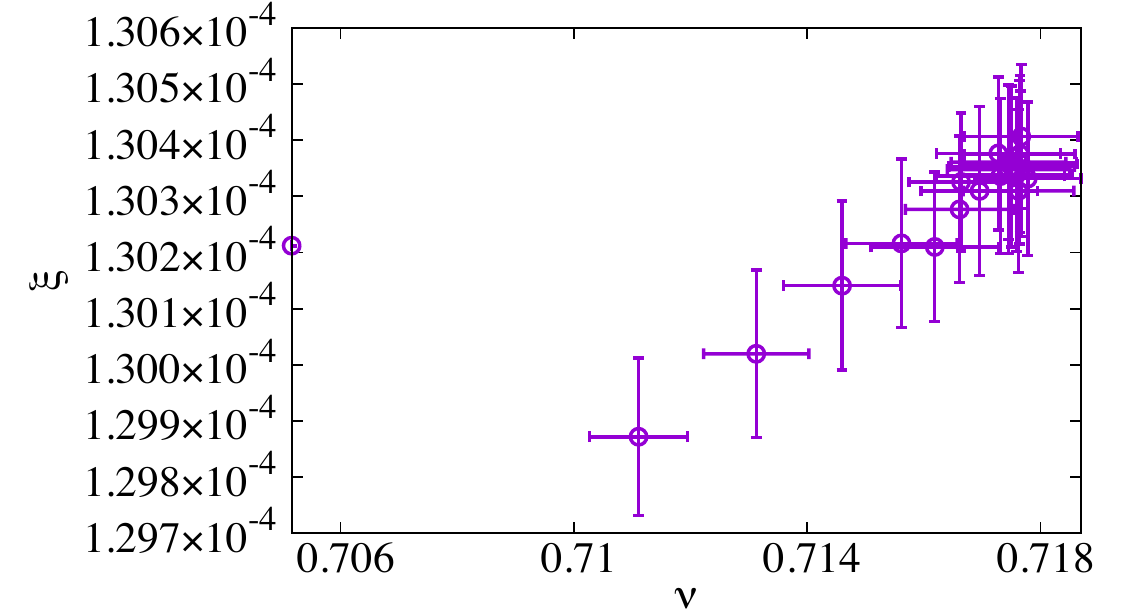} &
\includegraphics[width=0.4\textwidth]{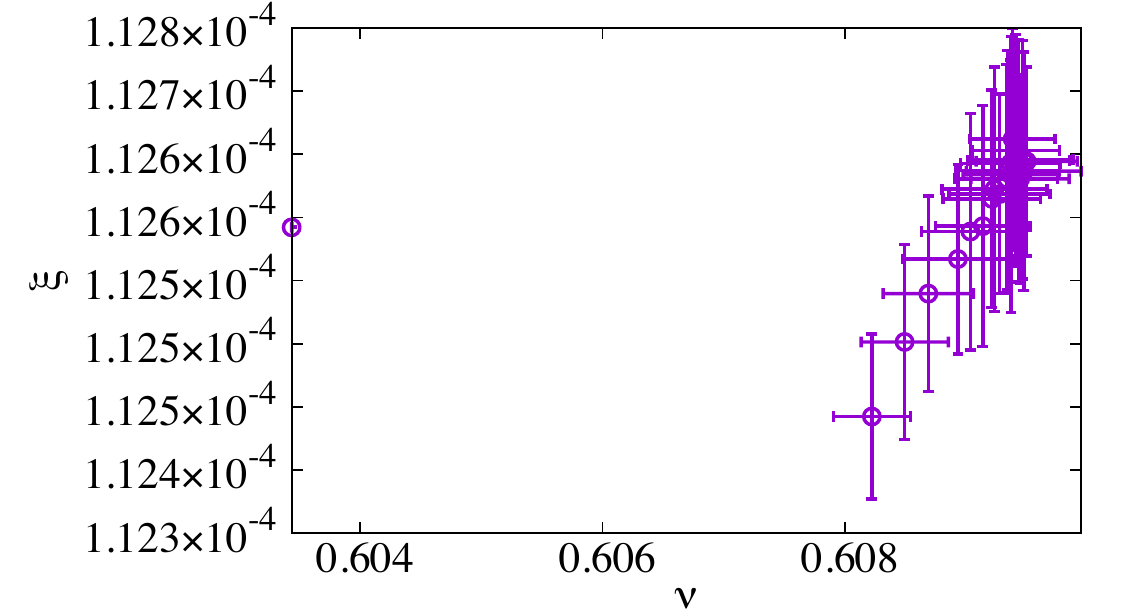} \\
\end{tabular}
\caption{Plot of $\nu$ in Eq.~\eqref{eq:NODF}~\cite{AlmeidaNeto2008} versus $\xi$ in Eq.~\eqref{NCQ_formula} for the synthetic network model with the 
same number of animals and plants (100) and the noise parameter (a) $\eta = 0.2$ and (b) $\eta=0.4$; the EPS process is applied later
with various $T$ Monte Carlo steps.
The leftmost points for both (a) and (b) correspond to $T = 0$.
The error bars in both the horizontal and vertical directions present the standard deviation
for each $\eta$ (100 network realizations for each $\eta$ and each $T$), except for $T = 0$ (a single ``original'' realization with $\eta=0.2$ and $0.4$, respectively).
}
\label{EPS_NODF_vs_NCQ_scatter_for_synthetic_networks}
\end{figure*}

Figure~\ref{NODF_vs_NCQ_scatter}(a) shows a strong correlation between the $\nu$ value~\cite{AlmeidaNeto2008} in Eq.~\eqref{eq:NODF} and the $\xi$ in Eq.~\eqref{NCQ_formula}, for the mutualistic network data introduced in Sec.~\ref{sec:data_sets},
which indeed supports the correspondence between the nestedness and core-periphery structure for this set of mutualistic networks.
However, for these data, in fact, the edge density (the number of edges divided by the maximum possible number of edges, i.e., the product of the numbers of animals and plants) is correlated with both $\nu$ and core quality similarly or even slightly stronger, as shown in
Figs.~\ref{NODF_vs_NCQ_scatter}(b) and \ref{NODF_vs_NCQ_scatter}(c). 
Therefore, we take the model introduced in Sec.~\ref{sec:synthetic_network_model} with $100$ animals, $100$ plants, and the noise parameter $\eta$ shown in 
Fig.~\ref{NODF_vs_NCQ_scatter_for_synthetic_networks}(a), and one can clearly see that the $\nu$ and $\xi$ values are extremely well correlated,
even if the numbers of nodes and edges are exactly the same (hence the edge density as well) for different $\eta$ by the model construction where the number 
of edge is strictly conserved during 
the edge reshuffling process.

However, even for this model where the average degree (or its first moment) is completely fixed, as shown in Figs.~\ref{NODF_vs_NCQ_scatter_for_synthetic_networks}(b)
and \ref{NODF_vs_NCQ_scatter_for_synthetic_networks}(c), the standard deviation (the second moment) of degree values is enough to distinguish the nestedness and
core-periphery-ness.  
To remove such an effect of degree distribution completely, we try model networks with the EPS process described in Sec.~\ref{sec:synthetic_network_model} to conserve the degree sequence.
As shown in Fig.~\ref{EPS_NODF_vs_NCQ_scatter_for_synthetic_networks}, although it is somewhat surprising that the EPS process actually increases the nestedness
(larger $\nu$ values), i.e., it \emph{induces} the nestedness instead of destroying it, except for the $T=0$ case (without the EPS process),
there is a positive correlation between nestedness and core-periphery-ness.

In other words, similar to the assortativity-clustering space of a network's degree sequence reported in Ref.~\cite{Holme2007},
the nestedness-coreness space seems to be quite restricted by the degree sequence.
There are recent studies on such restricted ensembles for edge shuffling of networks~\cite{Orsini2015,Fischer2015}.
More fundamentally, it is well known that not all degree distributions, or their actual realizations represented as degree sequences, can be assembled as resultant networks~\cite{DelGenio2011,YBaek2012}, so one can already see that there could be severe structural restriction on the realized networks.
In summary, as expected from the presumption, nestedness and core-periphery-ness are strongly correlated, but we cannot exclude the possibility that the underlying degree distribution itself may yield the resultant mesoscale structures. Further studies would be required to reveal more fundamental principles behind the connections. Indeed, the connection between nestedness and degree distributions has also been discussed~\cite{Bascompte2003,Burgos2007}, which we have to keep in mind, as we now know the severe restriction posed by the degree distribution.

\begin{figure*}
\begin{tabular}{lll}
(a) & (b) & (c) \\
\includegraphics[width=0.33\textwidth]{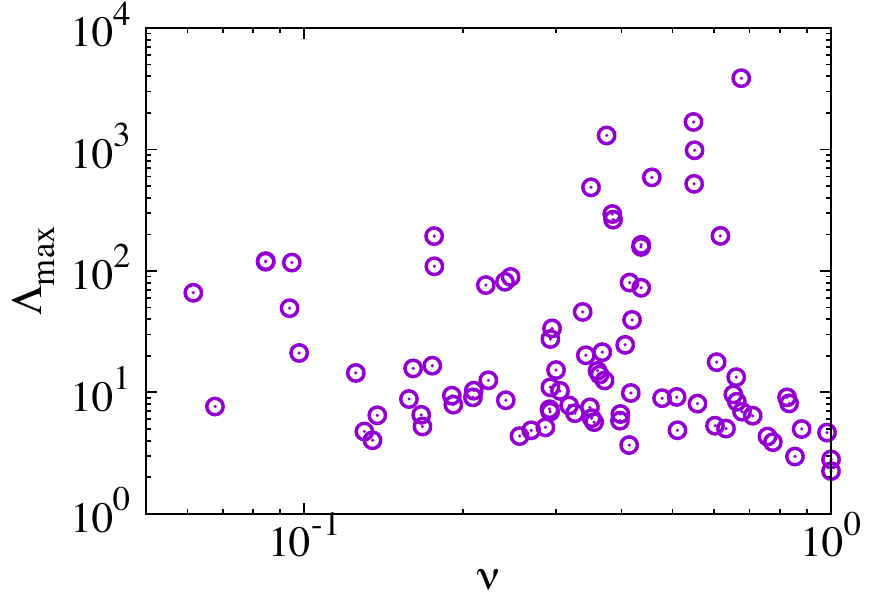} &
\includegraphics[width=0.33\textwidth]{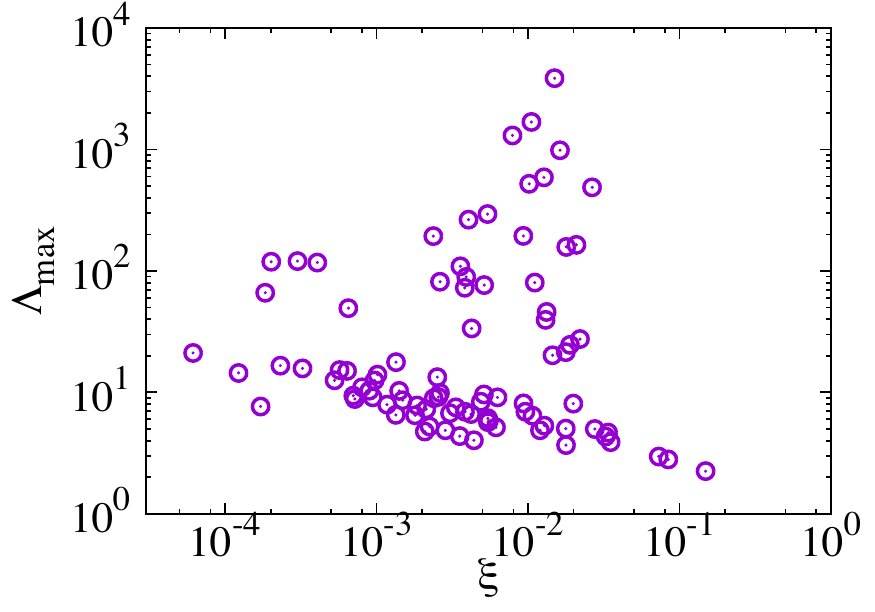} &
\includegraphics[width=0.33\textwidth]{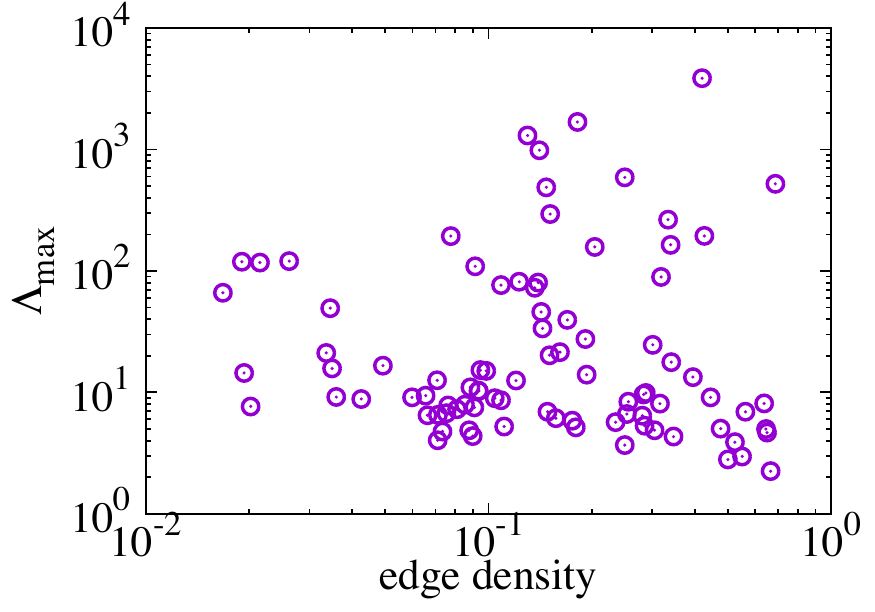} \\
\end{tabular}
\caption{For the mutualistic network data, (a) $\nu$ in Eq.~\eqref{eq:NODF} versus the largest eigenvalue $\Lambda_\mathrm{max}$, with the correlation coefficients $0.183$ with the $p$-value $= 0.196$ (Pearson), $-0.170$ with the $p$-value $= 0.112$ (Spearman), and $-0.114$ with the $p$-value $= 0.113$ (Kendall), (b) $\xi$ in Eq.~\eqref{NCQ_formula} versus the largest eigenvalue $\Lambda_\mathrm{max}$, with the correlation coefficients $0.020$ with the $p$-value $= 0.851$ (Pearson), $-0.142$ with the $p$-value $= 0.183$ (Spearman), and $-0.135$ with the $p$-value $= 0.061$ (Kendall), and (c) the edge density versus the largest eigenvalue $\Lambda_\mathrm{max}$, with the correlation coefficients $0.109$ with the $p$-value $= 0.311$ (Pearson), $-0.150$ with the $p$-value $= 0.162$ (Spearman), and a $-0.105$ with the $p$-value $= 0.145$ (Kendall).
}
\label{fig:largest_eigenvalue}
\end{figure*}

We have focused on the structural properties so far, but we have to consider the dynamical aspect of ecological networks as well. An aspect of dynamical stability on ecological networks can be assessed by the maximum eigenvalue of the adjacency matrix~\cite{Pastravanu2006}. Note that our definition of bipartite adjacency matrix $\{ W_{ij} \}$ in Sec.~\ref{sec:nestedness_and_c_p} is a nonsquare matrix in general, so we use the original definition of the adjacency matrix: $\{ A_{kl} \}$, where $A_{kl}$ represents the interaction between nodes $k$ and $l$, regardless of their identities as animals and plants, i.e., both $k$ and $l$ $\in \{ 1, \dots, N_\mathrm{animal} + N_\mathrm{plant} \}$, which guarantees the square symmetric matrix and real eigenvalues. To check if our nestedness and its deeply related coreness measure are related to the dynamical stability, we examine the interrelationship between $\nu$, $\xi$, and the edge density and the largest eigenvalue $\Lambda_\mathrm{max}$ of the adjacency matrix $\{ A_{kl} \}$ for the mutualistic network data as shown in Fig.~\ref{fig:largest_eigenvalue}. The result indicates that there is no statistically significant relation between our nestedness or coreness measures and the dynamical stability measured by the largest eigenvalue. In other words, the structural closeness of nestedness and core-periphery structures exists regardless of the dynamical stability, at least for this data set.

\section{Conclusions and Discussion}
\label{sec:discussion}

We have explored the seemingly obvious connection between the nestedness and core-periphery structure of networks, despite the fact that it looks obvious solely from the shape of adjacency matrix with proper ordering. We have shown the actual correlation between the two using a set of ecological mutualistic networks and model networks, by clearly addressing the nestedness and core-periphery-ness in the weighted and bipartite network level. In addition, in the process of generating null-model networks, we have found that given degree distributions set large restriction on both nestedness and core-periphery-ness, which hinders the investigation on the parameter space.

In any case, it is clear that the nestedness can be considered as a generalized (or finer) version of the core-periphery structure based on our observation. 
We may even consider other variants such as sorting the nodes with respect to the core-periphery-ness values instead of degree values to calculate the nestedness, e.g., modification of Eq.~\eqref{eq:NODF}, where the nodes are sorted based on $\xi$ in Eq.~\eqref{CS_formula} instead of degree.
Once we accept that those two concepts are closely related, albeit not equivalent, we can map many problems in regard to nestedness, such as its origin, and effects on the system of interest, to those of the core-periphery structures where we may be able to find answers more easily.

Finally, we would like to remark on the work on the nestedness and the community structure~\cite{CommunityReview} measured by the modularity function~\cite{Fortuna2010},
where the authors indeed found some degree of correlation between the two concepts. However, 
the correlations reported there are much weaker than the ones we report in this work. It is worth looking at the mesoscale 
properties of networks, but we believe that the core-periphery structure is the correct measure to compare, rather than the community structure.
The phrase ``two sides of the same coin'' included the title of Ref.~\cite{Fortuna2010} should be attached in fact to the nestedness versus core-periphery structure, not the community structure.

\begin{acknowledgments}

The author thanks Puck Rombach for the \texttt{MATLAB} code to calculate $\xi$ values and Young-Ho Eom (엄영호), Daniel Kim (김영호), and Petter Holme 
for fruitful discussions.

\end{acknowledgments}

\end{CJK}
\end{document}